\documentclass[3p,two column,fleqn]{elsarticle}
\biboptions{sort&compress}
\pdfoutput=1
\journal{Physics Letters B}
\usepackage{amsmath,physics,amssymb,graphics}
\usepackage{hyperref}
\hypersetup{pdfauthor=author}
\newcommand*\xbar[1]{%
	\hbox{%
		\vbox{%
			\hrule height 0.5pt % The actual bar
			\kern0.5ex%         % Distance between bar and symbol
			\hbox{%
				\kern-0.1em%      % Shortening on the left side
				\ensuremath{#1}%
				\kern0em
			}%
		}%
	}%
}

\begin{document}
\begin{frontmatter}
\title{Replica wormhole and AMPS firewall}
\author[mainaddress]{Amir A. Khodahami}
\ead{a.khodahami@shirazu.ac.ir}
\author[mainaddress]{Azizollah Azizi\corref{correspondingauthor}}
\ead{azizi@shirazu.ac.ir}
\address[mainaddress]{Department of Physics, Shiraz University, Shiraz 71949-84795, Iran}
\cortext[correspondingauthor]{Corresponding author}
\date{Received: date / Accepted: date}
\begin{abstract}
	We investigate the potential for generating an AMPS firewall from the replica wormhole topology, which describes the post-Page time spacetime in the evaporation process of a black hole. Our analysis reveals that this topology gives rise to a Dirac delta force experienced by infalling particles at the event horizon, consistent with the AMPS firewall hypothesis. This force, proportional to the particle's total energy squared, is a direct consequence of the inherent Dirac delta in the Ricci scalar of the replica wormhole. We provide a regularization scheme for this force, which is crucial for obtaining physically meaningful results. Notably, our approach suggests that considering the replica wormhole resolves both the information paradox and the AMPS argument regarding the monogamy of entanglement simultaneously, yielding a more coherent framework.
	\begin{keyword}
		AMPS Firewall \sep Replica Wormhole \sep Monogamy of Entanglement \sep Hawking Radiation.
	\end{keyword}
\end{abstract}
\end{frontmatter}
\raggedbottom
%%%%%%%%%%%%%%%%%%%%%%%%%%%%%%%%%%%%%%%%%%%%%%%%%%%%%%%%%%%%%%%%%%%%%%%%%%%%
\section{Introduction}
\begin{figure*}[t]
		\centering
			\includegraphics[width=0.95\textwidth]{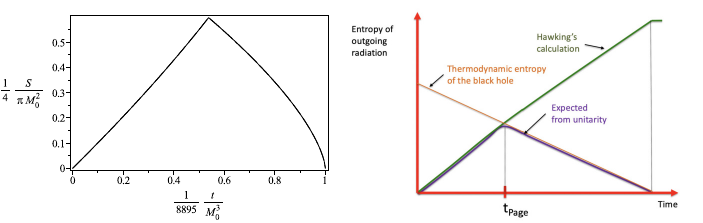}
		\caption{\label{f__Page} Left: The curve that Page suggested for entanglement entropy of the radiation emitted from a black hole initially in a pure state, from \cite{page2013a}. This curve is produced by considering the thermodynamic entropies of the radiation and the black hole as functions of time, and choosing the less one. Right: A schematic visualization of Page's procedure in obtaining the Page curve, from \cite{almheiri2021b}. As is shown, the true curve for entanglement entropy of the radiation is expected to follow the thermodynamic entropy of the radiation at first. However, when reaching the Page time, it should switch to follow the thermodynamic entropy of the black hole. Obviously this curve is consistent with unitarity.}
	\end{figure*}

In 1975, Hawking studied quantum field theory in curved spacetime and showed that black holes do evaporate while losing their masses \cite{hawking1975a,page1976a}. Given that the radiation is purely thermal, the evaporation process destroys quantum information and is not a unitary process \cite{hawking1976a}. A quantity called ``von Neumann entropy'' or ``fine-grained entropy'' or ``entanglement entropy'' is used to express the information loss amount quantitatively, which is defined as $S_R = -\Tr\rho_R\ln\rho_R$, with $\rho_R$ representing the density matrix of the radiation. One would expect this quantity to rise at the initial stages of the evaporation process due to the increasing entanglement between the emitting radiation and the remaining black hole. However, for a black hole starting in a pure state, the entanglement entropy should decrease after about half of the black hole has evaporated (Page time) and eventually hit zero for the process to be unitary. Hence, the entanglement entropy should follow the curve shown in Fig.~\ref{f__Page}, suggested by Page \cite{page1983b,page1993a,page2013a}, for the information to be conserved.
\par Further investigations has confirmed that the information is indeed conserved. In 2006, Ryu and Takayanagi introduced a holographic formula for the entanglement entropy, which can be proved in the context of AdS/CFT correspondence, using replica trick \cite{ryu2006c,fursaev2006,nishioka2009b}. In the derivation of the holographic formula we have
\begin{equation}\label{Eq-Dg}
	\Tr_A\rho_A^n = \int Dg \; \exp(-\mathcal{S}_{n\text{-AdS}}[g]),
\end{equation}
where $\rho_A$ is density matrix of a system $A$ described by a $d$-dimensional CFT. Moreover, the integration is performed over different geometrical configurations (given by different metric tensors) in the dual $n$-sheeted AdS space, which in turn is constructed by gluing $n$ numbers of $(d+1)$-dimensional AdS spaces by the means of \cite{nishioka2009b}, with $\mathcal{S}_{n\text{-AdS}}[g]$ being its action given by
\begin{equation}\label{Eq-S_nAdS}
	\mathcal{S}_{n\text{-AdS}}[g]=-\frac{1}{16\pi G}\int dx^{(d+1)}\sqrt{\left|\det(g)\right|}\left(R+\Lambda\right).
\end{equation}
$R$ in the above expression is the Ricci scalar of the $n$-AdS space and $\Lambda$ is the negative cosmological constant. The entanglement entropy, $S_A$, can be found using the following relation
\begin{equation}\label{Eq-Rep}
	S=-\frac{\partial}{\partial n}\ln(\Tr\rho^n)\biggl|_{n=1}.
\end{equation}
In the CFT side, there is a negative deficit angle of $2\pi(1-n)$ localized on the gluing surface, $\partial A$. Consequently, in the dual AdS space, the deficit angle locates on a codimension-2 surface homologous to $A$, we denote which by $\gamma_A$. The deficit angle contributes to the Ricci scalar through a singular Dirac delta, i.e.,
\begin{equation}\label{Eq-R}
	R=4\pi (1-n)\delta (\gamma_A)+R^{(0)},
\end{equation}
with $R^{(0)}$ representing the Ricci scalar of the AdS space and being extensive in $n$. $\gamma_A$ in the above relation is only restricted to be homologous to $A$ and hence is not quite fixed. So the summation over different geometries in Eq.~\eqref{Eq-Dg} breaks into: (i) a summation over different choices of $\gamma_A$ alongside (ii) a summation over different choices of $g$ in $R^{(0)}$. Using Eqs.~\eqref{Eq-Dg}-\eqref{Eq-R}, and approximating the summations over $g$ (in $R^{(0)}$) and $\gamma_A$ by their maximum contributions given by classical general relativity and minimal surface ($\gamma_A^*$) respectively, we obtain the holographic entanglement entropy formula
\begin{equation}\label{Eq-Hol}
	S_A=\frac{\text{Area}(\gamma_A^*)}{4G}.
\end{equation}
\par The holographic formula, Eq.~\eqref{Eq-Hol}, is particularly useful when we aim to compute the entanglement entropy of a system in a CFT. This formulation was further developed by \cite{hubeny2007b} to a covariant form and by \cite{faulkner2013c} to incorporate quantum corrections, eventually leading to the following formula proposed by \cite{engelhardt2015c}, which is applicable to evaporating black holes
	\begin{equation}\label{Eq__EE_QES}
		S=\text{min}_X\left\{\text{ext}_X\left[\frac{\text{Area}(X)}{4G}+S_{\text{semi-cl}}\left(\Sigma_X\right)\right]\right\}.
	\end{equation}
Here, $X$ is a codimension-2 surface and $\Sigma_X$ is the region bounded by $X$ and the surface from where we can almost assume that spacetime is flat, called cutoff surface\footnote{Determining the precise location of the cutoff surface is challenging according to the long-range nature of gravity \cite{almheiri2021b}.}. Moreover, the term $S_{\text{semi-cl}}(\Sigma_X)$ is standing for the entanglement entropy of the quantum fields on $\Sigma_X$ in the semi-classical description\footnote{The semi-classical description treats matter fields as quantum and the gravitational field as classical.}. The expression in the brackets is called \sl generalized entropy\rm, $S_\text{gen}(X)$. The surface that extremizes $S_\text{gen}(X)$ is called Quantum Extremal Surface (QES). According to the recipe by \cite{engelhardt2015c}, if there are multiple QESs, then we should select the one with the least $S_\text{gen}$.
\par For an evaporating black hole that starts from a pure state, we identify two QESs as described by Eq.~\eqref{Eq__EE_QES}: (i) a vanishing surface with a growing contribution, and (ii) a nonvanishing surface just behind the event horizon with a decreasing contribution. In the early stages, the minimal QES is the vanishing one causing the generalized entropy to match the entanglement entropy of the bulk (inside the cutoff surface). Hence, the initial black hole’s fine-grained entropy starts at zero and continuously increases. On the other hand, the nonvanishing QES appears shortly after the black hole formation with a contribution that reflects the black hole's area, which decreases as the black hole evaporates. According to the minimal selection rule of Eq.~\eqref{Eq__EE_QES}, when the contribution of the nonvanishing surface becomes smaller than that of the vanishing one, it starts to represent the true fine-grained entropy of the black hole \cite{almheiri2021b}. This switch between the two QESs allows the black hole's entanglement entropy to closely follow the Page curve shown in Fig.~\ref{f__Page}, which is indicative of unitary evaporation \cite{page1983b,page1993a,page2013a}.
\par
\begin{figure*}[t]
	\centering
	\includegraphics[width=0.8\textwidth]{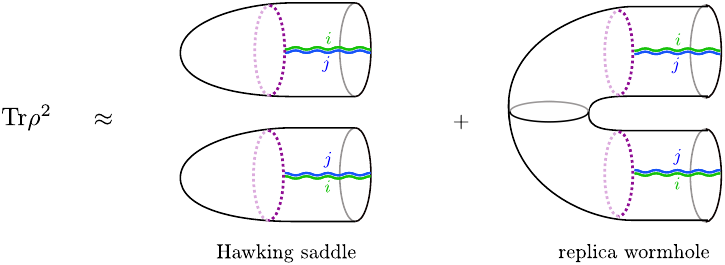}
	\caption{\label{f__HW} Illustrating diagrams are given in \cite{almheiri2021b} for two different contributions of $\Tr\rho^2$. Left is what Hawking considered in his calculations and hence is called Hawking saddle. It simply considers the black holes as evaporating independently and is the dominant contribution before the Page time. Right is a non-trivial topology in which the interiors of the black holes are connected forming a wormhole called replica wormhole. It becomes the dominant contribution after the Page time.}
\end{figure*}

 In the context of the replica trick, when assuming $n=2$ replicas, we have two different topologies contributing in $\Tr\rho^2$ as shown in Fig.~\ref{f__HW}. Notably, they are in accordance with the two minimal QESs discussed previously. The first diagram, called Hawking saddle, represents $n=2$ black holes evaporating independently. The second diagram, on the other hand, represents $n=2$ black holes in the way that their interiors are connected, forming a kind of wormhole, called replica wormhole. The minimization rule of Eq.~\eqref{Eq__EE_QES} leads to a strict switch between these saddles: Before the Page time, the Hawking saddle is dominant and after that time the replica wormhole becomes dominated. A more accurate treatment would replace the strict transition by a gradual one, and, correspondingly, smoothen the sharp peak of the Page curve \cite{khodahami2023}.
\par In 2012, Ahmed Almheiri, Donald Marolf, Joseph Polchinski, and James Sully (AMPS) revealed that the three fundamental assumptions in black hole physics: (i) purity of Hawking radiation, (ii) absence of infalling drama, and (iii) validity of semiclassical behavior outside the event horizon, are in conflict \cite{almheiri2013a}. This conflict arises because the post-Page time radiation has to be maximally entangled with both its ingoing pair (according to (iii)) and the pre-Page time radiation (enforced by (i)), violating the monogamy of entanglement in absence of an infalling drama (according to (ii)). To resolve this conflict, AMPS proposed that observers crossing the event horizon experience a ``firewall'' of high-energy particles, which must turn on prior to the emission of post-Page time radiation; i.e., at least around the Page time \cite{almheiri2013b}.‌ Although there are some alternative resolutions, such as complementarity \cite{susskind1993,muthukrishnan2023} and non-local interactions \cite{giddings2013a,giddings2013,giddings2023}, AMPS argued that removing assumption (ii) by introducing the firewall is the most conservative resolution to the paradox.
\par This investigation explores the potential for generating an AMPS firewall from the replica wormhole, through studying the Ricci scalar. The organization of this paper is as follows: In Sec.~\ref{Sec_Firewall}, learning from Eq.~\eqref{Eq-R}, we state that the Ricci scalar for the replica wormhole topology contains a singular Dirac delta multiplied by $(n-1)$. This combination becomes uncertain when we send $n\rightarrow 1$ and $r\rightarrow r_s$. Therefore, we keep this combination, while neglecting the $(n-1)$ alone according to the replica trick, to derive a metric that corresponds to the replica wormhole topology, expected to describe the post-Page time spacetime. Subsequently, we simulate an infalling particle toward the black hole in this spacetime and find that it experiences a Dirac delta force at the event horizon. This force is in accord with the AMPS firewall hypothesis. However, a regularization is needed to remove its ambiguity, which is explored in Sec.~\ref{Sec_n-1}. Finally conclusions are presented in Sec.~\ref{Sec_Conc}.
%%%%%%%%%%%%%%%%%%%%%%%%%%%%%%%%%%%%%%%%%%%%%%%%%%%%%%%%%%%%%%%%%%%%%%%%%%%%
\section{Firewall from replica wormhole}\label{Sec_Firewall}
For a Schwarzschild geometry, the Ricci scalar is zero as it inherently satisfies the equation $R=0$. Nevertheless, as a Schwarzschild black hole undergoes evaporation, two new elements emerge that can alter this null value: (i) the spacetime backreaction caused by the Hawking radiation, and (ii) the replica wormhole contribution appearing after the Page time. Assuming that the radiation is collected and transported to an asymptotic region allows us to conveniently disregard the backreaction of the Hawking radiation. To be more concrete, this assumption enables us to ignore the impact of the radiation on the regions of our interest, i.e., near the black hole. On the other hand, the replica wormhole should contribute as a Dirac delta multiplied by $(n-1)$, similar to Eq.~\eqref{Eq-R}. In light of our earlier discussions on the QESs, this Dirac delta is restricted to the event horizon of the black hole. Consequently, the Ricci scalar takes the form
\begin{equation}\label{Eq__R-new}
R=4\pi(1-n)\,\frac{k_1\,\delta(r-r_s)}{r_s} \,\theta(t-t_{\text{P}}),
\end{equation}
where $k_1$ is a dimensionless constant of order unity, $r_s$ denotes the Schwarzschild radius of the black hole, and $\theta(t-t_{\text{P}})$ is the Heaviside step function, equal to unity for times after the Page time $t_\text{P}$, and zero otherwise. Moreover, the denominator is added to make the Ricci scalar a true scalar density. As the Ricci scalar contains a Dirac delta multiplied by $(n-1)$, it becomes uncertain when we send $n\rightarrow 1$ and $r\rightarrow r_s$ simultaneously. We study the effect of such a combination on the preexisting Schwarzschild spacetime. Hence our limit of interest is: to ignore the terms with $(n-1)$ alone according to the replica trick, while considering them when they are multiplied by a Dirac delta; i.e., sending $(n-1)\rightarrow 0$ while keeping $(n-1)\delta(\cdots)$.
\par The unaltered Schwarzschild metric tensor reads as
\begin{equation}
	\begin{split}
		g^{(0)}_{\mu\nu}&=\\
		&\text{diag}\left\{1-r_s/r\,,\,-\frac{1}{1-r_s/r}\,,\,-r^2\,,\,-r^2\sin^2\theta\right\}.
	\end{split}
\end{equation}
Upon introducing the Ricci scalar $R$ in Eq.~\eqref{Eq__R-new}, the metric tensor undergoes a modification as \cite{grquick}
\begin{equation}\label{Eq__g-new}
	\begin{split}
		g_{\mu\nu}&=\\
		&\text{diag}\bigg\{(1-r_s/r)(1-a\theta(r-r_s)\theta(t-t_\text{P}))\,,\\
		&\qquad\qquad\quad-\frac{1}{1-r_s/r}\,,\,-r^2\,,\,-r^2\sin^2\theta\bigg\},
	\end{split}
\end{equation}
where $a=8\pi k_1(n-1)/3$. Specifically, the inclusion of $R$ affects the temporal-temporal component of the metric tensor by introducing a multiplication factor, i.e.,
\begin{equation}
g_{00}\rightarrow \left[1-a\theta(r-r_s)\theta(t-t_\text{P})\right]g_{00}.
\end{equation}
Interestingly, this alteration does not cause any change in the metric tensor within the limit of interest (sending $a\rightarrow 0$ while keeping $a\delta(\cdots)$). Still, it may impact the geodesic followed by an infalling particle toward the black hole, as the geodesic equations involve derivatives of the metric tensor. In this work, we simply focus on a radially infalling particle and neglect the angular motion. By adopting the perspective of the particle, the relevant geodesic equations (temporal and radial) can be expressed as \cite{grquick}
\begin{equation}\label{Eq__Geodesics-t}
	\begin{split}
	&t''(u)+\frac{r_s r'(u) t'(u)}{r(u)\left(r(u)-r_s\right)}\\
	&-a \biggl[r'(u)\delta\left(r(u)-r_s\right) \theta\left(t(u)-t_\text{p}\right)\\
	&+\frac{1}{2}t'(u)\delta\left(t(u)-t_\text{P}\right)\theta\left(r(u)-r_s\right)\biggr]t'(u)=0,
	\end{split}
\end{equation}
and
\begin{equation}\label{Eq__Geodesics-r}
	\begin{split}
	&r''(u)-\frac{r_s r'(u)^2}{2r(u)\left(r(u)-r_s\right)}+\frac{r_s\left(r(u)-r_s\right) t'(u)^2}{2 r(u)^3}\\
	&-a
	\biggl[r_s \theta\left(r(u)-r_s\right)+r(u)\left(r(u)-r_s\right)\delta\left(r(u)-r_s\right) \biggr]\\
	&\times\left(\frac{\left(r(u)-r_s\right)}{2 r(u)^3}\right)\theta\left(t(u)-t_\text{P}\right)t'(u)^2=0.
	\end{split}
\end{equation}
In the above equations, $r$ and $t$ represent the standard Schwarzschild coordinates, while $u$ denotes the particle's proper time. The prime sign signifies differentiation with respect to the proper time $u$. It is worth noting that these equations are derived under the intended limiting conditions (sending $a\rightarrow 0$ while keeping $a\delta(\cdots)$). Let focus on the radial equation (Eq.~\eqref{Eq__Geodesics-r}) as it governs the proper acceleration, which is equivalent to the force (per unit mass) experienced by the infalling particle. By introducing a finite mass for the particle, we can utilize the following definition for the proper time
\begin{equation}
	du^2:=ds^2=g_{00}dt^2+g_{11}dr^2,
\end{equation}
where $g_{00}$ and $g_{11}$ are given by Eq.~\eqref{Eq__g-new}. With this choice, together with the identity
\begin{equation}\label{Eq__r'}
	r'(u)=\frac{dr}{du}=\frac{dr}{dt}\cdot \frac{dt}{du}=v(u)t'(u),
\end{equation}
we have
\begin{equation}\label{Eq__t'}
	t'(u)=\left(g_{00}+v(u)^2g_{11}\right)^{-1/2}.
\end{equation}
The letter $v$ in the above equations stands for the Schwarzschild velocity, $dr/dt$. Inserting from Eqs.~\eqref{Eq__r'} and \eqref{Eq__t'} into the radial geodesic equation, Eq.~\eqref{Eq__Geodesics-r}, and applying the intended limit, we obtain \cite{grquick}
\begin{equation}\label{Eq__r''}
	r''=-\frac{r_s}{2r^2}+a\left(\frac{(r-r_s)^3\delta\left(r-r_s\right) }{2r\left((r-r_s)^2-r^2v^2\right)}\right)\theta\left(t-t_\text{P}\right),
\end{equation}
where we have omitted the $u$-dependence for sake of brevity. The first term on the right-hand side of the above equation represents the familiar gravitational force (per unit mass) experienced by the infalling particle. However, the second term, appearing after the Page time, highlights the impact of the replica wormhole on the particle's geodesic. To evaluate this expression, we need to substitute in the Schwarzschild velocity $v$, which for a geodesic particle is given by
\begin{equation}\label{Eq__v}
	v=\left(1-\frac{r_s}{r}\right)\left(1-c^{-1}\left(1-\frac{r_s}{r}\right)\right)^{1/2},
\end{equation}
where $c$ denotes the square of the total energy per mass of the particle ($c=E^2/m^2$), which should not be confused with the speed of light (since we are working in natural units)\footnote{As a reminder, in a Schwarzschild spacetime, the total energy of a particle is a constant of motion, given by
\begin{equation}
	E=\frac{m\sqrt{1-r_s/r}}{\sqrt{1-w^2}},
\end{equation}
where
\begin{equation}
	w=\frac{\sqrt{-g_{11}}dr}{\sqrt{g_{00}}dt}=\frac{v}{(1-r_s/r)}.
\end{equation}
}. Substituting $v$ from Eq.~\eqref{Eq__v} into Eq.~\eqref{Eq__r''}, we arrive at \cite{grquick}
\begin{equation}
	r''=-\frac{r_s}{2r^2}+\frac{E^2}{2m^2}\,a\delta\left(r-r_s\right)\,\theta\left(t-t_\text{P}\right).
\end{equation}
The second term on the right-hand side of the equation now explicitly illustrates the effect of the replica wormhole on the geodesic of a radially infalling particle with a total energy per mass of $E/m$. This term is describing a strict wall, thanks to the unbounded Dirac delta (and the positive coefficients), located at the event horizon. Moreover, it emerges after the Page time as intended. Hence, in accord with the AMPS firewall, it can be interpreted as ``firewall force'' (per unit mass) experienced by the infalling particle. However, the precise strength of this force remains unclear due to the uncertain combination $a\delta(r-r_s)$. We will address this issue in the following section by proposing a regularization for it. Notably, this force is proportional to the total energy squared of the particle, indicating that its strength depends on the particle's energy.
\par Before closing this section, we would like to conclude that deriving the firewall force from the replica wormhole provides a solution that resolves both the information paradox and the AMPS argument regarding the monogamy of entanglement. In fact, considering the replica wormhole topology as the first correction to Hawking's calculations offers a unified approach that simultaneously addresses both paradoxes, yielding a more coherent framework.
%%%%%%%%%%%%%%%%%%%%%%%%%%%%%%%%%%%%%%%%%%%%%%%%%%%%%%%%%%%%%%%%%%%%%%%%%%%%
\section{A regularization for the firewall force}\label{Sec_n-1}
In this section, we seek a regularization for the firewall force (per mass)
\begin{equation}\label{Eq__firewall}
	\frac{8\pi k_1}{3}\cdot\frac{E^2}{2m^2}\,(n-1)\delta\left(r-r_s\right)\,\theta\left(t-t_\text{P}\right).
\end{equation}
To achieve this, one might use the Gaussian representation of the Dirac delta
\begin{equation}
	\delta(r-r_s)=\lim_{l\rightarrow 0}\,\frac{1}{\sqrt{2\pi}\,l}\exp(-\frac{(r-r_s)^2}{2l^2}).
\end{equation}
A more accurate treatment should replace the limit $l\rightarrow0$ with $l\rightarrow l_\text{Pl}$, where $l_\text{Pl}$ represents the Planck length, which serves as the fundamental length scale. This substitution leads to a regularization for the Dirac delta
\begin{equation}\label{Eq__reg-delta}
	\delta(r-r_s)\rightarrow\frac{1}{\sqrt{2\pi}\,l_\text{Pl}}\exp(-\frac{(r-r_s)^2}{2l_\text{Pl}^2}).
\end{equation}
Therefore, the resultant firewall force will be proportional to the Planck mass $m_\text{Pl}= 1/l_\text{Pl}$. It is noteworthy that the expression converges to the initial Dirac delta as $l_\text{Pl}\rightarrow0$, indicating that the unboundedness of the Dirac delta is regulated by the finite value of $l_\text{Pl}$. As the term $(n-1)$ arises from the replica trick, it should be regarded as a mathematical quantity independent of the physical variables in the problem. As a result, we expect the firewall force magnitude to remain at $m_\text{Pl}$. Given $r_s$ and $l_\text{Pl}$ as two inherent length scales in this problem, maintaining the firewall force strength at $m_\text{Pl}$ necessitates that $(n-1)$ is of $\mathcal{O}((l_\text{Pl}/r_s)^0)$. On the other hand, regularizing the firewall force requires regularizing $(n-1)$ by $l_\text{Pl}$ as well, ensuring that $(n-1)$ vanishes as $l_\text{Pl}\rightarrow0$. In conclusion, the only option remaining is logarithmic regularization, i.e.,
\begin{equation}\label{Eq__reg-n-1}
	(n-1)\rightarrow\frac{k_2}{\ln\left(r_s/l_\text{Pl}\right)},
\end{equation}
with $k_2$ being a dimensionless constant of order unity. From Eqs.~\eqref{Eq__reg-delta} and \eqref{Eq__reg-n-1}, the uncertain firewall force (per mass), Eq.~\eqref{Eq__firewall}, is regularized as
\begin{equation}
	\frac{4\pi k_3}{3\sqrt{2\pi}}\cdot\frac{E^2}{m^2}\cdot \frac{\theta\left(t-t_\text{P}\right)}{l_\text{Pl}\ln\left(r_s/l_\text{Pl}\right)}\exp(-\frac{(r-r_s)^2}{2l_\text{Pl}^2}),
\end{equation}
with $k_3=k_1k_2$, a dimensionless constant of order unity.
%%%%%%%%%%%%%%%%%%%%%%%%%%%%%%%%%%%%%%%%%%%%%%%%%%%%%%%%%%%%%%%%%%%%%%%%%%%%
\section{Conclusions}\label{Sec_Conc}
In the evaporation process of a black hole, the replica wormhole topology emerges as the dominant contribution in the corresponding path integral after the Page time. Notably, its associated Ricci scalar features a Dirac delta term, $(n-1)\delta(r-r_s)$, which has motivated us to investigate the possibility of generating a firewall from it. To explore this further, we developed a metric that describes the post-Page time spacetime and calculated the geodesic equations for a particle falling radially toward the black hole. Our analysis revealed that the particle experiences a Dirac delta force at the event horizon, in line with the AMPS firewall argument. Notably, this approach indicates that taking into account the replica wormhole addresses both the information paradox and the AMPS argument concerning the monogamy of entanglement simultaneously, resulting in a more cohesive framework. However, the coefficient $(n-1)$ resulting from the replica trick should ideally approach zero. Consequently, as the particle approaches the event horizon ($r \rightarrow r_s$), this specific combination becomes uncertain, necessitating regularization. Our chosen approach involves scaling the resulting firewall force to the Planck mass.
\end{document}